\documentstyle[onecolumn,epsfig]{mn}
\newif\ifAMStwofonts
\ifoldfss
  \ifCUPmtlplainloaded \else
    \NewTextAlphabet{textbfit} {cmbxti10} {}
    \NewTextAlphabet{textbfss} {cmssbx10} {}
    \NewMathAlphabet{mathbfit} {cmbxti10} {} 
    \NewMathAlphabet{mathbfss} {cmssbx10} {} 
  \fi
  \ifAMStwofonts
    \ifCUPmtlplainloaded \else
      \NewSymbolFont{upmath} {eurm10}
      \NewSymbolFont{AMSa} {msam10}
      \NewMathSymbol{\upi}     {0}{upmath}{19}
      \NewMathSymbol{\umu}     {0}{upmath}{16}
      \NewMathSymbol{\upartial}{0}{upmath}{40}
      \NewMathSymbol{\leqslant}{3}{AMSa}{36}
      \NewMathSymbol{\geqslant}{3}{AMSa}{3E}

       \let\le=\leqslant
       \let\ge=\geqslant
    \fi
  \fi
\fi 

\ifnfssone
  \newmathalphabet{\mathit}
  \addtoversion{normal}{\mathit}{cmr}{m}{it}
  \addtoversion{bold}{\mathit}{cmr}{bx}{it}
  \newmathalphabet{\mathbfit} 
  \addtoversion{normal}{\mathbfit}{cmr}{bx}{it}
  \addtoversion{bold}{\mathbfit}{cmr}{bx}{it}
  \newmathalphabet{\mathbfss} 
  \addtoversion{normal}{\mathbfss}{cmss}{bx}{n}
  \addtoversion{bold}{\mathbfss}{cmss}{bx}{n}
  \ifAMStwofonts
    \ifCUPmtlplainloaded \else
      \UseAMStwoboldmath
      \makeatletter
      \new@mathgroup\upmath@group
      \define@mathgroup\mv@normal\upmath@group{eur}{m}{n}
      \define@mathgroup\mv@bold\upmath@group{eur}{b}{n}
      \edef\UPM{\hexnumber\upmath@group}
      \new@mathgroup\amsa@group
      \define@mathgroup\mv@normal\amsa@group{msa}{m}{n}
      \define@mathgroup\mv@bold\amsa@group{msa}{m}{n}
      \edef\AMSa{\hexnumber\amsa@group}
      \makeatother
      \mathchardef\upi="0\UPM19
      \mathchardef\umu="0\UPM16
      \mathchardef\upartial="0\UPM40
      \mathchardef\leqslant="3\AMSa36
      \mathchardef\geqslant="3\AMSa3E

       \let\le=\leqslant
       \let\ge=\geqslant
    \fi
  \fi
\fi 

\ifnfsstwo
  \DeclareMathAlphabet{\mathbfit}{OT1}{cmr}{bx}{it}
  \SetMathAlphabet\mathbfit{bold}{OT1}{cmr}{bx}{it}
  \DeclareMathAlphabet{\mathbfss}{OT1}{cmss}{bx}{n}
  \SetMathAlphabet\mathbfss{bold}{OT1}{cmss}{bx}{n}
  \ifAMStwofonts
    \ifCUPmtlplainloaded \else
      \DeclareSymbolFont{UPM}{U}{eur}{m}{n}
      \SetSymbolFont{UPM}{bold}{U}{eur}{b}{n}
      \DeclareSymbolFont{AMSa}{U}{msa}{m}{n}
      \DeclareMathSymbol{\upi}{0}{UPM}{"19}
      \DeclareMathSymbol{\umu}{0}{UPM}{"16}
      \DeclareMathSymbol{\upartial}{0}{UPM}{"40}
      \DeclareMathSymbol{\leqslant}{3}{AMSa}{"36}
      \DeclareMathSymbol{\geqslant}{3}{AMSa}{"3E}

       \let\le=\leqslant
       \let\ge=\geqslant
    \fi
  \fi
\fi 

\ifCUPmtlplainloaded \else
  \ifAMStwofonts \else 
    \def\upi{\pi}
    \def\umu{\mu}
    \def\upartial{\partial}
  \fi
\fi

\title[Resonant capture, counter-rotating disks, and polar rings]{Resonant capture, counter-rotating disks, and polar rings}
\author[S. Tremaine \& Q. Yu]
       {Scott Tremaine and Qingjuan Yu\\
Princeton University Observatory, Peyton Hall,Princeton, NJ~08544-1001, USA}
\date{Accepted ???
      Received ???
      in original form ???}

\pagerange{\pageref{firstpage}--\pageref{lastpage}}
\pubyear{2000}

\def\ltsima{$\; \buildrel < \over \sim \;$}
\def\simlt{\lower.5ex\hbox{\ltsima}}
\def\gtsima{$\; \buildrel > \over \sim \;$}
\def\simgt{\lower.5ex\hbox{\gtsima}}
\def\be{\begin{equation}}
\def\ee{\end{equation}}
\def\kms{{\rm\,km\,s^{-1}}}
\def\kpc{{\rm\,kpc}}

\def\yr{{\rm\,yr}}

\def\Gyr{{\rm\,Gyr}}

\def\half{{\textstyle{1\over2}}}

\def\bfr{{\bf r}}

\def\bfw{{\bf w}}
\def\p{\partial}

\def\omit#1{}
\def\p{\partial}

\def\bfr{{\bf r}}
\def\bfk{{\bf k}}

\def\bfI{{\bf I}}
\def\bfOmega{{\bf \Omega}}

\begin{document}
\maketitle

\label{firstpage}

\begin{abstract}
We suggest that polar rings and/or counter-rotating disks in flattened
galaxies can be formed from stars captured at the Binney resonance, where the
rate of precession of the angular momentum vector of a disk star equals the
pattern speed of a triaxial halo. If the halo pattern speed is initially
retrograde and slowly decays to zero, stars can be trapped as the Binney
resonance sweeps past them, and levitated into polar orbits. If the halo pattern
speed is initially retrograde and slowly changes to prograde, trapped stars
can evolve from prograde to retrograde disk orbits. The stellar
components of polar rings formed by this process should consist of two equal,
counter-rotating star streams.
\end{abstract}


\section{Introduction}

\noindent
This paper is motivated by the remarkable discovery \cite{rub92}
that the otherwise normal E7/S0 galaxy NGC 4550 contains two stellar disks
rotating in opposite directions.\nocite{rub92} The two disks are similar in
total luminosity and scale length and approximately coplanar \cite{rix92};
one is accompanied by an extended gas disk. 

A second case of a counter-rotating disk is the Sab galaxy NGC 7217, in which
20-30\% of the disk stars are on retrograde orbits, independent of radius
\cite{mk94}. Counter-rotating disks are rare: Kuijken, Fisher \& Merrifield
(1996) \nocite{kfm96}examined 28 S0 galaxies and found no counter-rotating
components containing more than $\sim5\%$ of the total disk light. In
contrast, roughly 20\% of the gas disks in S0 galaxies counter-rotate
\cite{ber92}; however, these gas disks are generally much smaller than the
accompanying stellar disks, so even if they form stars (as in NGC 3593,
Bertola et al. 1996) they are unlikely to produce two stellar disks of similar size
as in NGC 4550 or NGC 7217.

Several formation mechanisms for counter-rotating disks have been
discussed. (i) A merger could add material on retrograde orbits to a
pre-existing stellar disk, but mergers are likely to overheat the original
disk even in favourable cases where the merging galaxy is gas-rich or its
orbit lies in the original disk plane \cite{ryd98}. (ii) Hierarchical models
of galaxy formation predict that the mean angular momentum of infalling gas
varies substantially during the lifetime of a galaxy, so that early infall
could produce a gas disk that later forms stars, while late infall
subsequently brings in retrograde gas that forms stars in turn. This proposal
does not explain why the scale lengths of counter-rotating disks are similar
\cite{ryd98}, or why NGC 4550 satisfies the normal Tully-Fisher relation
\cite{rix92}. (iii) Evans \& Collett (1994)\nocite{ec94} point out that box
orbits in a triaxial potential can evolve into loop orbits if the potential
slowly becomes more axisymmetric (an effect expected from late infall;
Dubinski 1994). After this process stars will occupy both direct and retrograde
loops---in precisely equal numbers if the triaxial potential is
non-rotating---thereby naturally creating counter-rotating disks\footnote{The
dynamical stability of counter-rotating disks is discussed by Sellwood \&
Merritt \nocite{sm94}(1994).}.

Polar-ring galaxies are early-type (usually S0) galaxies containing an outer
ring of gas, dust and stars on orbits that are approximately circular and
nearly perpendicular to the symmetry plane of the galaxy. At least 0.5\% of S0
galaxies have polar rings, although this is probably an underestimate because
of orientation-dependent selection effects \cite{whi90}. The catalog by
Whitmore et al. lists only six kinematically confirmed polar-ring galaxies but
many more candidates without kinematic data. Like counter-rotating disks,
polar rings are usually assumed to form from the merger of a companion galaxy
or late gas infall (e.g. Katz \& Rix 1992, Bekki 1998); if the gravitational potential of
the primary galaxy is triaxial then there is a range of initial conditions
from which dissipative material will settle into a polar orbit
perpendicular to the long axis \cite{scd84,tvp94}.

In this paper we describe a novel way to form counter-rotating disks and/or
polar rings. The inclinations of disk-star orbits can be excited by resonant
coupling to a triaxial halo potential. The location of the relevant resonances
is determined by the vertical and azimuthal frequencies, $\Omega_2$ and
$\Omega_3$ (see \S \ref{sec:action}), and the pattern speed of the halo,
$\Omega_p$. In particular, Binney (1978, 1981)\nocite{bin78,bin81} has stressed
the importance of the resonance at
\be 
\Omega_3-\Omega_2=\Omega_p,
\label{eq:binney}
\ee
which we call the Binney resonance. The Binney resonance occurs when the
pattern speed matches the rate of precession of the angular momentum vector
or  node, $\dot\omega=\Omega_3-\Omega_2$. For low-inclination orbits in 
oblate potentials, $\Omega_2>\Omega_3$, so the Binney resonance
occurs for retrograde pattern speed.

We suppose that the disk is embedded in a triaxial halo that initially rotates
with a retrograde pattern speed $\Omega_{pi}<0$. The pattern speed is expected
to change slowly as the halo acquires dark matter by infall; we assume that
the pattern speed increases, reaching a final value $\Omega_{pf}\ge0$. If
$\Omega_p$ changes sufficiently slowly, stars with small initial inclinations
$i_0$ are trapped in the Binney resonance as $\Omega_p$ sweeps past
$\Omega_3-\Omega_2$. As $\Omega_p$ increases further, the orbits of the
trapped stars are levitated \cite{st96} to higher inclination while
remaining nearly circular, becoming polar orbits as $\Omega_p$ crosses
zero. Thus if $|\Omega_p|$ gradually decays to zero, an outer polar ring is
formed from disk stars with $\Omega_2-\Omega_3<|\Omega_{pi}|$.  If $\Omega_p$
crosses zero to positive values, the trapped stars evolve onto retrograde
orbits, until they are eventually released from the Binney resonance when
$\Omega_p$ sweeps through $-\Omega_2+\Omega_3>0$. The inclinations after
release are near $\pi-i_0$, so the flipped stars form a counter-rotating disk
with the same thickness as the initial disk.

A close cousin of this process was already discussed by\nocite{hei82} Heisler,
Merritt \& Schwarzschild (1982; see also van Albada, Kotanyi \& Schwarzschild 1982). They recognized that a
sequence of ``anomalous'' inclined orbits bifurcated from the closed
short-axis loop orbits at the Binney resonance in a rotating triaxial
potential, and speculated that gas might evolve onto the anomalous orbits as a
result of dissipation. However, they focused on a sequence of orbits of
decreasing energy at fixed pattern speed, which terminates in a short-axis
orbit that cannot be occupied by collisional material. In contrast, we examine
the behavior of dissipationless material in a system with time-varying pattern
speed. Our mechanism is also related to the levitation process discussed by
Sridhar \& Touma \nocite{st96}(1996), who focus on a different resonance as a
means of forming a thick disk.

In \S \ref{sec:action} we describe the classification of resonances
in nearly axisymmetric potentials and justify our focus on the Binney
resonance. In \S \ref{sec:hoop} we investigate capture and release of stars at
the Binney resonance using a simplified dynamical model. \S \ref{sec:num}
describes the results of numerical integrations, and \S \ref{sec:disc}
contains a discussion.

\section{Resonance classification}

\label{sec:action}

\noindent
To determine the most promising sites for resonant capture, we first consider
integrable motion in an axisymmetric potential. We define action-angle
variables $(\bfI,\bfw)$, such that $H_0=H_0(\bfI)$ is the axisymmetric
Hamiltonian and the trajectory of a particle is given by
\be 
\bfI=\hbox{const}, \qquad \bfw=\hbox{const}+\bfOmega t, \ee 
where 
\be 
\bfOmega(\bfI)={\p H_0\over \p \bfI}.  
\ee

The actions in a spherical potential can be chosen as follows: $I_1$ is the
radial action, which is zero for circular orbits; $I_2$ is the vertical or
latitudinal action, which is zero for prograde equatorial orbits and in
general is equal to $J-J_z$, where $J$ is the total angular momentum and $J_z$
its $z$-component; $I_3$ is the azimuthal action, which is equal to
$J_z$. Note that $I_1\ge0$, $I_2\ge0$, $I_3\le\half I_2$. 

The geometrical interpretation of these actions remains similar for integrable
axisymmetric Hamiltonians \cite{dZ85}; in particular, $I_3$ is still equal
to $J_z$, prograde equatorial orbits still have $I_2=0$, and the analogs of
circular orbits ($I_1=0$) are shell orbits, which occupy a two-dimensional
axisymmetric surface of zero thickness.

Now we add a weak, non-axisymmetric perturbing potential of the form
\be 
U_1(\bfr,t)=A(r,\theta)\cos\{m[\phi-\phi_p(t)]\},
\ee
where $m\not=0$ is a positive integer and $\dot\phi_p\equiv \Omega_p$ is the
pattern speed. In action-angle variables this potential can be written as 
(e.g. Tremaine \& Weinberg 1984) 
\be
U_1(\bfI,\bfw,t)=\sum_{\bfk} A_\bfk(\bfI)\cos[\bfk\cdot\bfw-m\phi_p(t)]
\label{eq:actang}
\ee
where $\bfk$ is an integer triple, with $k_3=m$.  Terms in the perturbing
potential (\ref{eq:actang}) with $k_3=0$ conserve the azimuthal action
$I_3=J_z$; terms with $k_1=0$ conserve the radial action $I_1$; and terms with
$k_2=0$ conserve the vertical action $I_2=J-J_z$. When $I_1=0$ the radial
angle $w_1$ is 
undefined, so $A_\bfk(I_1=0)$ must vanish for $k_1\not=0$; more generally
$A_\bfk\to I_1^{|k_1|/2}$ as $I_1\to 0$. Similarly $A_\bfk\to I_2^{|k_2|/2}$
as $I_2\to 0$ (in celestial mechanics these constraints are called
the d'Alembert characteristic). We shall assume that the perturbing potential
is symmetric around the equatorial plane, which requires that $A_\bfk=0$ unless
$k_2$ is even. 

We are interested in the resonant excitation of inclination in
low-inclination, near-circular disk orbits. Resonance occurs when the rate of
change of the resonant angle $\bfk\cdot\bfw-m\phi_p(t)$ is zero, yielding the
resonance condition 
\be 
\bfk\cdot\bfOmega=m\Omega_p.
\label{eq:rescond}
\ee
Since $A_\bfk\to 0$ as $I_1\to 0$ when $k_1\not=0$, the strongest inclination
excitation for near-circular orbits comes from terms with
$k_1=0$. We shall also focus on 
terms with $k_3=m=2$, since barlike and triaxial perturbations are the
strongest non-axisymmetric features in galaxies. Since $A_\bfk\propto
I_2^{|k_2|/2}$, near-equatorial orbits are most strongly affected by terms
with $k_2=0,\pm 2$ (recall that $k_2$ must be even); however, resonances 
with $k_2=0$ cannot excite inclination (for these resonances $w_2$ is
ignorable, so the vertical 
action $I_2$ is conserved), so we restrict our attention to $k_2=\pm2$. 
For the triples $\bfk=(0,\pm 2,2)$ the resonance condition (\ref{eq:rescond}) 
is
\be
\pm\Omega_2+\Omega_3=\Omega_p.
\ee
In potentials that are not highly flattened, the vertical and azimuthal
frequencies are similar, $\Omega_2\sim \Omega_3$. If the pattern speed is slow,
$|\Omega_p|<\Omega_{2,3}$ (as would be expected for the perturbation from a
triaxial halo) resonance is therefore more likely to occur for $k_2=-2$ than
for $k_2=2$. Thus we are led naturally to the Binney resonance
$\bfk=(0,-2,2)$ (eq. \ref{eq:binney}). 

At the Binney resonance $I_1$ is adiabatically invariant because the
perturbations with $k_1\not=0$ are rapidly varying. Moreover $w_2$ and $w_3$
appear in the potential only in the combination $2(w_3-w_2)$, so $I_2+I_3$ is
conserved. In nearly spherical potentials $I_2+I_3\simeq J$, so the total
angular momentum is approximately conserved. In other words the resonance
affects only the inclination of the orbit, and not its total angular momentum
or eccentricity. In effect, the evolution of a circular orbit near the Binney
resonance is that of a rigid spinning hoop with the same radius and
angular momentum, an analogy that we shall pursue in the next section.

\section{A simplified model for resonance capture}

\label{sec:hoop}

\noindent
We consider motion in a rotating, triaxial potential of the form
\be
U(\bfr,t)=\Phi\left(\tilde x^2+{\tilde y^2\over p^2} + 
{z^2\over q^2}\right);
\ee
here the rotating coordinates $(\tilde x,\tilde y)$ are related to the
inertial coordinates $(x,y)$ by
\be
x+iy=(\tilde x+i\tilde y)\exp(i\phi_p),\qquad {d\phi_p\over dt}=\Omega_p(t),
\label{eq:rotdef}
\ee
where $\Omega_p$ is the pattern speed. We assume that the $x$-axis is the
long axis, so $0<p,q\le 1$, and that the symmetry plane of the disk
is the $x$-$y$ plane. Since loop orbits around the intermediate axis
are unstable, the $z$-axis must be the short axis, so $q<p$. Thus
\be
0<q<p\le 1.
\label{eq:tri}
\ee
We assume that the axis ratios are not too far from unity---typically we
choose $q=0.8$, $p=0.9$---so we can expand the potential as
\be
U(\bfr,t)=\Phi(r^2)+{V^2\over 2r^2}\left[(p^{-2}-1)\tilde
y^2+(q^{-2}-1)z^2\right];
\ee
here we have replaced $d\Phi/dr^2$ by $\half V^2/r^2$ where $V(r)$ is the
circular speed at radius $r$.

For simplicity we assume that the
time variation of the pattern speed is given by
\be
\Omega_p(t)=\dot\Omega_pt,
\label{eq:omdot}
\ee
where $\dot\Omega_p$ is a constant. 

Since we are interested in nearly circular orbits, we replace the particle by
a circular hoop of radius $r$; the angular momentum of the hoop is a constant
$J$ and its orientation is specified by the canonical momentum-coordinate pair
$(J_z,\omega)$, where $J_z=J\cos i$ is the $z$-component of the angular
momentum, $i$ is the inclination, and $\omega$ is the longitude of the
ascending node. The Hamiltonian of the hoop is $H(J_z,\omega,t)=\langle
U(\bfr,t)\rangle$, where $\langle\cdot\rangle$ denotes an average over the
hoop. Neglecting unimportant constants we have 
\begin{eqnarray} 
H(J_z,\omega,t) & = &{V^2\over 4}\left\{(p^{-2}-1)[\sin^2(\omega-\phi_p)
+\cos^2i\cos^2(\omega-\phi_p)]+(q^{-2}-1)\sin^2i\right\} \nonumber \\
& = & -{V^2\over 8}\left[(2q^{-2}-p^{-2}-1) {J_z^2\over J^2} +
(p^{-2}-1)\left(1-{J_z^2\over J^2}\right)\cos2(\omega-\phi_p)\right].
\label{eq:hamtoy}
\end{eqnarray}
Throughout this analysis we can consider $V$ to be constant, since the radius
of the hoop is fixed. 

We now convert to new canonical variables $(W,w)=(J_z/J,\omega-\phi_p)$ and a
dimensionless time $s=t/(J\beta)$, where $\beta$ is a constant to be chosen
below. Using equation (\ref{eq:omdot}) for the time-dependence of the pattern
speed, the new Hamiltonian is found to be 
\begin{eqnarray}
H(W,w,s)& = &-{\beta V^2\over 8}(2q^{-2}-p^{-2}-1)W^2 
\nonumber \\ & & \qquad 
-{\beta V^2\over 8}(p^{-2}-1)(1-W^2)\cos 2w-(\beta^2J^2\dot\Omega_p)Ws.
\end{eqnarray}
We choose the timescale parameter $\beta$ so that the coefficient of
$W^2$ is $-\half$; thus
\be
H(W,w,s)=-\half W^2 - \alpha(1-W^2)\cos 2w - \gamma Ws,
\label{eq:ham}
\ee
where
\begin{eqnarray}
\alpha &=&{p^{-2}-1\over 2(2q^{-2}-p^{-2}-1)}, \nonumber \\
\beta & = & {4\over V^2(2q^{-2}-p^{-2}-1)}, \nonumber \\
\gamma & = & {16r^2\dot\Omega_p\over V^2(2q^{-2}-p^{-2}-1)^2};
\end{eqnarray}
here we have replaced $J$ by $rV$, its value when $p$ and $q$ are near unity. 
The constraints (\ref{eq:tri}) imply that 
\be
0<\alpha<\half, \qquad \beta>0,
\label{eq:ablim}
\ee 
and $\hbox{sgn}(\gamma)=\hbox{sgn}(\dot\Omega_p)$; for the
axis ratios $q=0.8$, $p=0.9$, we have $\alpha=0.1317$. 

The Hamiltonian (\ref{eq:ham}) depends on two parameters: $\alpha$, which
determines the strength of the non-axisymmetric potential, and $\gamma$, which
determines the speed of the resonance passage. In the limit where the
potential is nearly spherical, $1-p,1-q=\hbox{O}(\epsilon)\ll 1$, we have
$\alpha=\hbox{O}(1)$, $\beta=\hbox{O}(\epsilon^{-1})$,
$\gamma=\hbox{O}(\dot\Omega_p\epsilon^{-2})$. The phase space is
a sphere of unit radius with a simple physical interpretation: the
angular-momentum vector {\bf J} has azimuth $w+\phi_p-\half\pi$ and colatitude
$i=\cos^{-1}W$.

We are interested in the case where the pattern speed changes slowly,
$|\dot\Omega_p|\ll \epsilon^2$ or $|\gamma|\ll 1$. In this case, over short
times the trajectory closely follows a level curve of the Hamiltonian for
fixed $s$, $H(W,w,s)=h(s)$ (the ``guiding trajectory''). The nature of the
guiding trajectories depends on the topology of the level surfaces of $H$. In
describing this topology we restrict ourselves to $0<\alpha<\half$, as
required by equation (\ref{eq:ablim}); we may then distinguish the following
stages (Figure \ref{fig:aitoff}):

\begin{figure}
\centerline{\psfig{figure=fig1.epsi,width=0.4\textwidth}}
\caption{Topology of the level surfaces of the Hamiltonian (\ref{eq:ham}), for 
$\alpha<\half$. The plots show equal-area
Aitoff projections of the sphere with longitude $w$ and colatitude
$i=\cos^{-1}W$, where $i$ is the inclination and $w\in[-\pi,\pi]$ is the 
longitude of the ascending node in the frame rotating at the pattern speed 
$\Omega_p$. The
plots show only stages (a)--(c); stages (d) and (e) are upside-down versions
of (b) and (a) respectively.}
\label{fig:aitoff}
\end{figure}

\begin{description}

\item[(a)] $\gamma s < -2\alpha -1$: There are stable equilibrium points at
$W=1$ (north pole) and $W=-1$ (south pole); the angle $w$ circulates for all
orbits.

\item[(b)] $-2\alpha-1< \gamma s < 2\alpha -1$: There is a stable equilibrium
point at the south pole, and an unstable equilibrium or saddle point at the
north pole. There are also stable equilibria at $w=\pm\half\pi$, $W=-\gamma
s/(1+2\alpha)$. The separatrix orbit passing through the north pole has energy
$h_{\rm sep}=-\half-\gamma s$, and divides circulating orbits with $h<h_{\rm
sep}$ from librating orbits with $h>h_{\rm sep}$.

\item[(c)] $2\alpha -1 < \gamma s < -2\alpha +1$: There are stable equilibrium
points at both poles, as well as stable equilibria at $w=\pm\half\pi$,
$W=-\gamma s/(1+2\alpha)$. In addition there are unstable equilibria at
$w=\{0,\pi\}$, $W=-\gamma s/(1-2\alpha)$. The separatrix orbit passing through
the unstable equilibria has energy $h_{\rm sep}=\half(\gamma
s)^2/(1-2\alpha)-\alpha$, and divides circulating orbits with $h<h_{\rm sep}$
from librating orbits with $h>h_{\rm sep}$.

\item[(d)] $-2\alpha + 1 < \gamma s < 2\alpha +1$: This is identical to stage
(b) after the transformation $\gamma s\to-\gamma s$, $W\to -W$. There are
stable and unstable equilibria at the north and south poles respectively, and
stable equilibria at $w=\pm\half\pi$, $W=-\gamma s/(1+2\alpha)$.

\item[(e)] $2\alpha + 1 < \gamma s$: This is identical to stage (a) after the
transformation $\gamma s\to-\gamma s$, $W\to -W$: there are stable equilibria
at $W=\pm1$ and $w$ circulates for all orbits.

\end{description}

Adiabatic invariance ensures that over long times the guiding trajectory
evolves so as to preserve the action, which is $(2\pi)^{-1}$ times the area on
the phase-space sphere enclosed by the guiding trajectory (e.g
Peale 1976, Henrard 1982, Borderies \& Goldreich 1984, Engels \& Henrard 1994).

Let us follow the evolution of the orbit for the case $\gamma>0$,
corresponding to a pattern speed that is initially negative (retrograde) but
increasing. At large negative time, $\gamma s \ll -1$, the Hamiltonian
(\ref{eq:ham}) is dominated by the term $-\gamma Ws$ and the guiding
trajectory is $W=W_0=$constant, where $W_0=\cos i_0$ and $i_0$ is the initial
inclination. The initial action is $(2\pi)^{-1}$ times the
area of the north polar cap on the phase-space sphere that is enclosed by the
initial trajectory, and is equal to $1-W_0$. 

As the time $s$ increases, the topology of the Hamiltonian eventually changes
from stage (a) to stage (b). In the initial phases of stage (b), the guiding
trajectory continues to circulate, despite the growing libration zones defined
by the separatrix orbit through the north pole. Eventually the area occupied
by the libration zones grows to $2\pi(1-W_0)$, so the action can no longer be
conserved if the orbit continues to circulate. At this point the guiding
trajectory crosses the separatrix and is captured into one of
the two libration zones.

The area occupied by the libration zones in stage (b) can be evaluated
analytically,
\be
A(\gamma s)={4\gamma s\over (1-4\alpha^2)^{1/2}}\cos^{-1}{1+\gamma
s-4\alpha^2\over 2\alpha\gamma s} +4\cos^{-1}{-1-\gamma s\over 2\alpha},
\qquad -2\alpha-1\le\gamma s\le 2\alpha-1.
\ee
As the time $s$ increases, $A(\gamma s)$ increases from zero at the onset of
stage (b) to $A_{\rm max}$ at the onset of stage (c), where
\be
A_{\rm max}=A(2\alpha-1)=
4\pi\left[1-\left(1-2\alpha\over 1+2\alpha\right)^{1/2}\right].
\ee
Capture into libration occurs when $A=2\pi(1-W_0)$
and is certain if  $A_{\rm max}>2\pi(1-W_0)$; in other words capture is certain
if  
\be 
\cos i_0 > 2\left(1-2\alpha\over 1+2\alpha\right)^{1/2}-1. 
\label{eq:inc} 
\ee 
For our nominal value $\alpha=0.1317$, capture is certain if the initial
inclination $i_0<58.2^\circ$. For larger inclinations, capture is
probabilistic because the guiding trajectory crosses the separatrix orbit
during stage (c), where transition to either libration or circulation can
occur. In this case the capture probability can be computed using methods
described by Henrard (1982). 

The captured orbits remain librating through stage (c) and in the initial
phases of stage (d). Eventually they re-cross the separatrix into circulation;
by symmetry their final inclination is just $i_f=\pi-i_0$. In other words, a
disk of stars in direct rotation is flipped into a disk with the same
radial profile and thickness but in retrograde rotation. This mechanism
operates if (i) The initial pattern speed $\Omega_{pi}$ corresponds to stage
(a) and the final pattern speed $\Omega_{pf}$ to stage (e); this requires 
\be
\Omega_{pi}<-{V\over 2r}(q^{-2}-1),\qquad\Omega_{pf}> {V\over 2r}(q^{-2}-1).
\label{eq:resconda}
\ee 
This condition can be restated in terms of $\dot\omega=\Omega_3-\Omega_2$,
the nodal precession rate for low-inclination orbits, as
$\Omega_{pi}<\dot\omega$, $\Omega_{pf}>-\dot\omega$.  (ii) The pattern speed
changes slowly enough that the action is adiabatically invariant except near
the separatrix. (iii) The initial inclination of the disk
stars is sufficiently small (eq. \ref{eq:inc}).

Another interesting outcome occurs if the pattern speed is initially
retrograde and slowly decays to zero. In this case the stars will be trapped
in polar orbits (stage (c) with $s=0$); the trapping process populates the two
separatrices equally, so the resulting polar ring will itself form a
counter-rotating disk, perpendicular to the long axis of the triaxial
potential.

\section{Numerical results}

\label{sec:num}

\noindent
We have followed the evolution of test-particle orbits in a rotating triaxial
potential of the form
\be
U(\bfr,t)=\half V^2\log\left(\tilde x^2+{\tilde y^2\over p^2} + 
{z^2\over q^2}\right),
\label{eq:trilog}
\ee
where $V$ is a constant and $\tilde x$ and $\tilde y$ are defined by 
equation (\ref{eq:rotdef}). We
take axis ratios $q=0.8$, $p=0.9$ and set $V=1$. The particle is initially on
a circular orbit with radius $r=1$. The pattern speed is assumed to vary as
\be
\Omega_p(t)=\Omega_{pi}+(\Omega_{pi}-\Omega_{pf})\left(e^{-t/\tau}-1\right),
\label{eq:pattern}
\ee
and the orbits are followed from $t=0$ to $t=4\tau$. 
With these parameters the star can undergo resonant capture and release if
(eq. \ref{eq:resconda})
\be
\Omega_{pi}<-0.281, \qquad \Omega_{pf}>0.281.
\ee
\begin{figure}
\centerline{\psfig{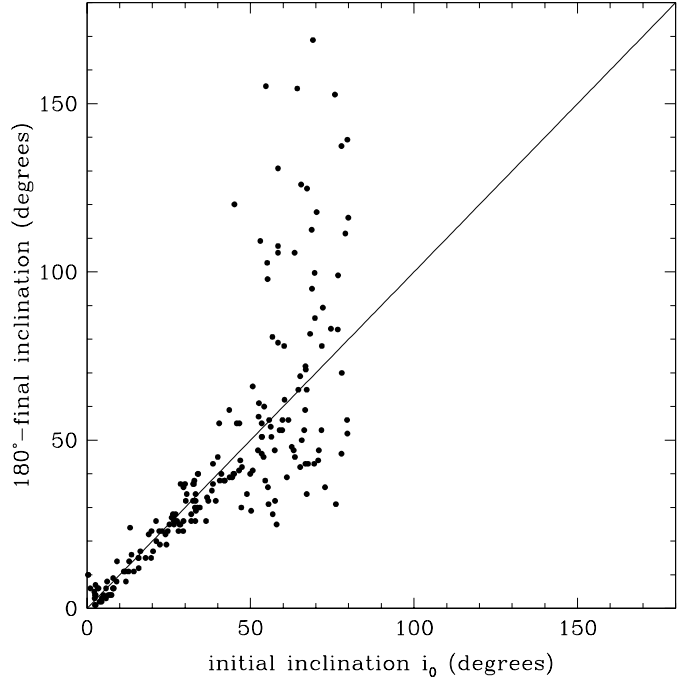}}
\caption{
Initial inclination $i_0$ versus final inclination for initially
circular orbits with radius $r=1$ in the triaxial potential defined by
equation (\ref{eq:trilog}). The evolution of the pattern speed is given by
equation (\ref{eq:pattern}), with $\Omega_{pf}=-\Omega_{pi}=0.4$ 
and $\tau=5000$.For $i_0\simlt 50^\circ$, 
the final inclination is $\simeq180^\circ-i_0$. 
}
\label{fig:two}
\end{figure}

\begin{figure}
\centerline{\psfig{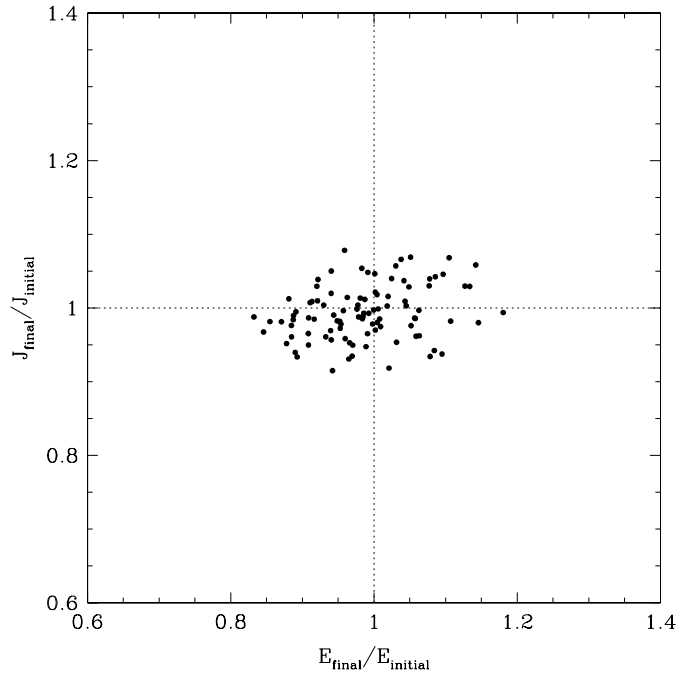}}
\caption{Final energy and angular momentum for the flipped particles in Figure
\ref{fig:two} ($i_0<40^\circ$). The initial energy and angular momentum are
shown by dashed lines. The changes in both quantities are small, confirming
that the size and shape of the orbit are not changed by the Binney
resonance.}
\label{fig:three}
\end{figure}

To illustrate resonant capture and release in the adiabatic limit, we have
numerically integrated the orbits of $200$ test particles with random orbital
phases and nodes, varying the pattern speed according to the parameters
$\Omega_{pi}=-0.4$, $\Omega_{pf}=+0.4$, $\tau=5000$. The results are shown in
Figure \ref{fig:two}: stars with initial inclination $i_0\la 50^\circ$
are flipped to inclination $\sim 180^\circ-i_0$, while for $i_0\ga
50$--$60^\circ$ the final inclination exhibits a wide spread. These results
are consistent with the conclusion from \S \ref{sec:hoop} that resonant
capture was certain for these parameters if $i_0<58.2^\circ$. Figure
\ref{fig:three} shows that the fractional changes in total angular momentum
and energy of the flipped particles are $\la 0.1$, confirming that
capture in the Binney resonance changes only the inclination of the particle
orbits, not their size or shape.

\begin{figure}
\centerline{\psfig{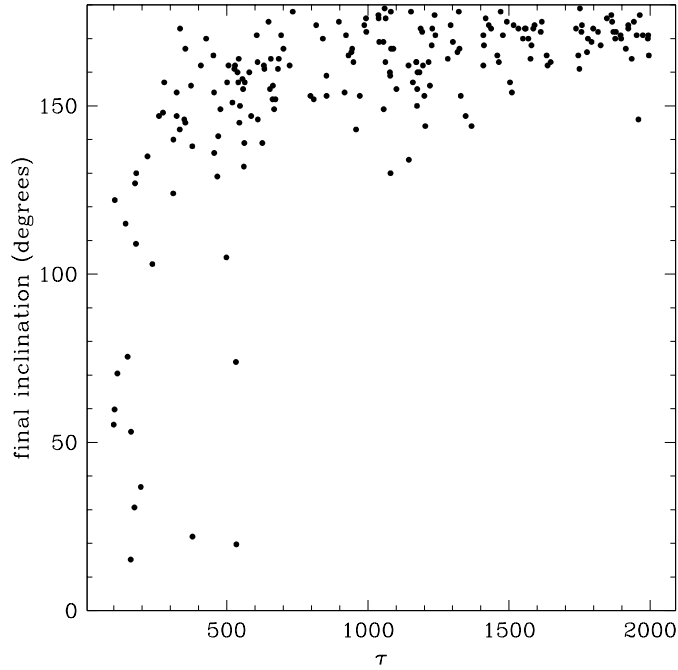}}
\caption{(top) Final inclination for initially circular orbits with radius
$r=1$ in the triaxial potential defined by equations (\ref{eq:trilog}) and
(\ref{eq:pattern}), with $\Omega_{pi}=-0.4$, $\Omega_{pf}=0.4$, and $\tau$
varying from 100 to $2000$. The initial inclinations are distributed as
$dN\propto i_0\exp(-\half i_0^2/\sigma^2)di_0$, where $\sigma=5^\circ$. The 
final inclinations are clustered near $180^\circ$ for $\tau\simgt 300$. }
\label{fig:gauss}
\end{figure}

To investigate the validity of the adiabatic approximation, we have integrated
200 test particles with $\tau$ varying from 100 to 2000, and initial
inclinations distributed as $n(i_0)di_0\propto i_0di_0\exp(-\half
i_0^2/\sigma^2)$, with $\sigma=5^\circ$. The distribution of final
inclinations shown in Figure \ref{fig:gauss} shows that most stars are
captured when $\tau\ga 300$; in physical units this corresponds to
$\tau \ga 7.5\Gyr (r/5\kpc)(200\kms/V)$. 

\begin{figure}
\centerline{\psfig{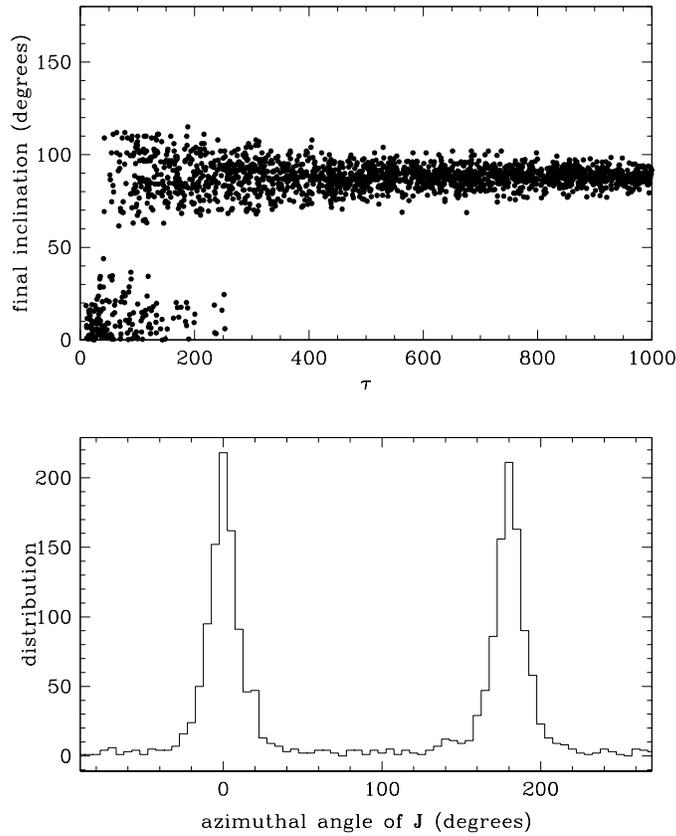}}
\caption{(top) Final inclination for initially
circular orbits with radius $r=1$ in the triaxial potential defined by
equations (\ref{eq:trilog}) and (\ref{eq:pattern}), with
$\Omega_{pi}=-0.4$, $\Omega_{pf}=0$, and $\tau$ varying from 10 to $10^3$. The
final inclinations are clustered near $90^\circ$, so the captured stars form a
polar ring. (bottom) Distribution of the azimuthal angle of the final angular
momentum vector in the rotating frame. The angular momentum vectors are centered on the long or
$x$-axis of the potential, with equal numbers of stars rotating in opposite
directions.} 
\label{fig:four}
\end{figure}

We have also investigated the case where the pattern speed is initially
retrograde and slowly decays to zero. Figure \ref{fig:four} shows the
distribution of final inclinations as a function of the timescale $\tau$, for
$\Omega_{pi}=-0.4$, $\Omega_{pf}=0$. The initial inclination range was
$i_0=0$--$10^\circ$. For $\tau\ga 100$ most of the particles are trapped
in orbits with inclination near $90^\circ$, thus forming a polar ring. The
angular-momentum vectors are aligned with the long axis of the potential, with
equal numbers of stars having $J_x>0$ and $J_x<0$. Thus the polar ring is
itself a counter-rotating disk.

Finally, we have examined whether resonance capture can occur in disk
galaxies, by integrating test-particle orbits in a non-axisymmetric 
Miyamoto-Nagai potential 
\be
\Phi(R,z)=-{GM\over\left(\tilde x^2+\tilde y^2/ p^2
+(a+(b^2+z^2)^{1/2})^2\right)^{1/2}}, 
\label{eq:mn}
\ee
where $\tilde x$ and $\tilde y$ are defined in equation (\ref{eq:rotdef}) and
$a=0.5$, $b=0.1$. The particles are initially in circular orbits of unit
radius, and the pattern speed varies according to (\ref{eq:pattern}) with
$\Omega_{pi}=-2$, $\Omega_{pf}=2$. We find that for initial inclination
$i_0\la 6^\circ$, most particles are captured
and flipped into retrograde orbits with inclination $180^\circ-i_0$. The
maximum inclination of the flipped orbits depends on the parameter $b$, which
controls the thickness of the disk mass distribution; for $b=0.2$ most
particles are flipped if $i_0\la 15$--$20^\circ$. 

\begin{figure}
\centerline{\psfig{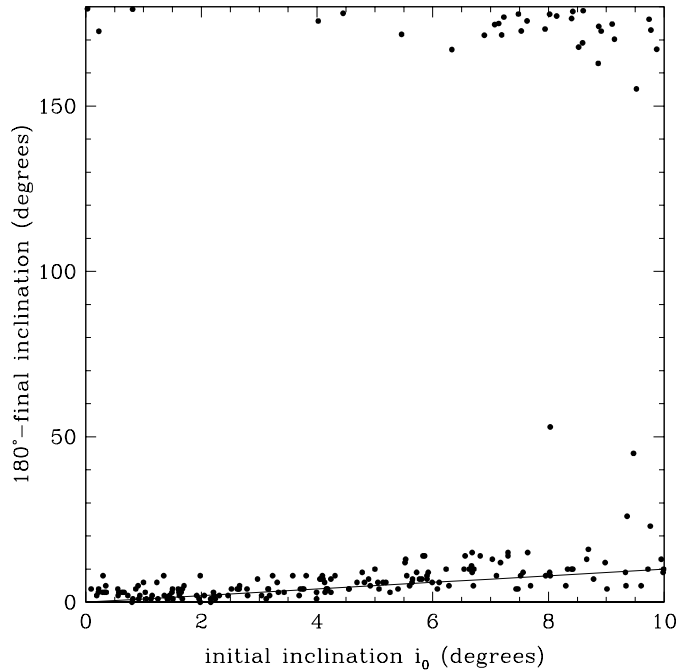}}
\caption{Initial inclination $i_0$ versus final inclination for initially
circular orbits with radius $r=1$ in the Miyamoto-Nagai potential
(\ref{eq:mn}) with $a=0.5$, $b=0.1$. The evolution of the pattern speed is
given by equation (\ref{eq:pattern}), with
$\Omega_{pf}=-\Omega_{pi}=2$ and $\tau=2000$. For $i_0\simlt 6^\circ$, the
final orbits are retrograde and equatorial.}
\label{fig:six}
\end{figure}

\section{Discussion}

\label{sec:disc}

\noindent
In rotating triaxial potentials with time-varying pattern speed, stars on
short-axis loop orbits can be captured at the Binney resonance, where the
pattern speed matches the precession rate $\dot\omega$ of the angular momentum
vector. As the pattern speed continues to change, captured stars will be
carried to high inclinations without large changes in energy or total angular
momentum.

Binney's (1981)\nocite{bin81} original discussion focused on the linear
response of stars close to the resonance when the pattern speed was fixed. In
contrast, by exploiting the theory of slow (adiabatic) resonant capture
\cite{pea76,hen82,bg84,eh94}, we can follow the nonlinear orbital evolution
so long as the variation in pattern speed is slow enough.

It is striking that slow resonant capture and subsequent inclination growth is
certain for stars with inclination $i\simeq0$, no matter how weak the triaxial
potential may be, even though $i=0$ is a formal solution of the equations of
motion for the model Hamiltonian (\ref{eq:hamtoy}) for all time. The
resolution of this apparent paradox is that (i) a Mathieu-type linear
inclination instability is always present when the star is sufficiently close
to resonance (Binney 1981); (ii) as the strength of the non-axisymmetry
approaches zero, the drift rate of the pattern speed must also approach zero
in order that the adiabatic approximation is valid.

If the final pattern speed of the triaxial potential is near zero, stars
captured into the Binney resonance will form a polar ring of long-axis loop
orbits. This model explains naturally why polar rings do not extend to the
center of the galaxy: only stars whose initial precession rate $|\dot\omega|$
is less than the initial pattern speed $|\Omega_{pi}|$ can be captured. The
stellar component of polar rings formed by this mechanism should exhibit two
equal, counter-rotating star stream, a testable prediction. 

If the final pattern speed is positive and larger than the precession rate of
the angular momentum vector, $\Omega_{pf}>|\dot\omega|$, stars captured into
the resonance with initial inclination $i_0$ will be flipped onto retrograde
orbits and then released with final inclination $\pi-i_0$. In disk potentials
only the low-inclination fraction of the disk orbits are captured and flipped
(Fig. \ref{fig:six}), so both the direct and retrograde disk can be composed
of pre-existing stars. For mildly triaxial potentials, capture is certain for
orbits with small or moderate inclinations; in this case we must rely on
subsequent star formation to re-form the parent disk.  This model explains
naturally why the two components of the counter-rotating disks in NGC 4550 and
NGC 7217 have similar scale lengths. It does not explain why the two
components in NGC 4550 have similar luminosity but then the two components in
NGC 7217 do not, and in any case there are strong selection effects that favor
the discovery of counter-rotating disks with similar luminosity.

A concern with this model is whether the required timescale
for variation of the pattern speed is unrealistically slow. The time unit in
our simulations is $2.4\times10^7 \yr\-(r/5\kpc)\-(200\kms/V)$, so timescales
$\tau \ga 400$ may exceed $10^{10}\yr$, the natural timescale for
variations in halo pattern speed. This should be compared with $\tau\ga
300$ required to flip orbits in the triaxial potential (Fig. \ref{fig:gauss})
and $\tau\ga 1000$ to flip orbits in the Miyamoto-Nagai disk
(Fig. \ref{fig:six}).  On the other hand we have used over-simplified model
potentials and have not explored parameter space systematically.

Another issue is whether gravitational noise due to molecular clouds or spiral
arms degrades the effectiveness of resonant capture, although this is not a
problem for the majority of polar rings that are found in S0 galaxies. 

An unresolved question is whether significant quantities of gas can be
captured into the Binney resonance. The orbits of particles in the two
libration zones around $w=\pm\half\pi$ intersect, so gas clouds in the two
zones will collide at high speed, leading to rapid energy dissipation. There
are thus two possibilities: either no gas is captured, or all of the gas is
captured into one of the libration zones, leaving the other vacant. Numerical
simulations are the best way to determine which of these two outcomes is more
realistic\footnote{Numerical simulations of polar rings
\cite{hi85,var90,qui91,kr92,ckrh92,bek98} show that gas rings on the
long-axis loop orbits in triaxial potentials can be stable in the presence of
dissipation, even though the libration zone surrounds a local maximum of the
averaged Hamiltonian (\ref{eq:hamtoy}).}. 

This research was supported in part by NSF Grant AST-9900316 and NASA Grant
NAG5-7066.

\end{document}


\end{document}